\documentclass[conference,a4paper]{IEEEtran}
\usepackage[utf8]{inputenc} 
\usepackage[T1]{fontenc}
\usepackage{url}              
\usepackage{cite}             

\usepackage[cmex10]{amsmath}  

\usepackage{mleftright}       

\usepackage{cgitIEEE}		  
\mleftright

\usepackage{graphicx}         
\usepackage{booktabs}         

\usepackage{algorithmicx}    

\usepackage[caption=false,font=footnotesize]{subfig}

\IEEEoverridecommandlockouts

\begin{document}

\title{Analysis of the Relative Entropy Asymmetry in the Regularization of Empirical Risk Minimization
\thanks{This work is supported by the University of Sheffield ACSE PGR scholarships, the Inria Exploratory Action -- Information and Decision Making (AEx IDEM), and in part by a grant from the C3.ai Digital Transformation Institute.}
}


%
%
%
 \author{%
   \IEEEauthorblockN{Francisco Daunas\IEEEauthorrefmark{1}\IEEEauthorrefmark{2},
                     I\~naki Esnaola\IEEEauthorrefmark{1}\IEEEauthorrefmark{3},
                     Samir M.~Perlaza\IEEEauthorrefmark{2}\IEEEauthorrefmark{3}\IEEEauthorrefmark{4},
                     and H.~Vincent Poor\IEEEauthorrefmark{3}}
   \IEEEauthorblockA{\IEEEauthorrefmark{1}%
                     ACSE Dept. University of Sheffield,
                     Sheffield, United Kingdom.
                     \{jdaunastorres1, esnaola\}@sheffield.ac.uk}
   \IEEEauthorblockA{\IEEEauthorrefmark{2}%
                     INRIA, 
                     Centre Inria d'Universit\'e C\^ote d'Azur,
                     Sophia Antipolis, France.
                     samir.perlaza@inria.fr}
   \IEEEauthorblockA{\IEEEauthorrefmark{3}%
                     ECE Dept. Princeton University, Princeton, 
                     08544 NJ, USA.
                     poor@princeton.edu}
   \IEEEauthorblockA{\IEEEauthorrefmark{4}%
                     GAATI, Universit\'e de la Polyn\'esie Fran\c{c}aise,
                     Faaa, French Polynesia.}
 }

\maketitle

\begin{abstract}
  %
  The effect of the relative entropy asymmetry is analyzed in the empirical risk minimization with relative entropy regularization (ERM-RER) problem. 
  A novel regularization is introduced, coined Type-II regularization, that allows for solutions to the ERM-RER problem with a support that extends outside the support of the reference measure.
  The solution to the new ERM-RER Type-II problem is analytically characterized in terms of the Radon-Nikodym derivative of the reference measure with respect to the solution.
  The analysis of the solution unveils the following properties of relative entropy when it acts as a regularizer in the ERM-RER problem:
  $i\bigl)$ relative entropy forces the support of the Type-II solution to collapse into the support of the reference measure, which introduces a strong inductive bias that dominates the evidence provided by the training data;
  $ii\bigl)$ Type-II regularization is equivalent to classical relative entropy regularization with an appropriate transformation of the empirical risk function. 
Closed-form expressions of the expected empirical risk as a function of the regularization parameters are provided.


\end{abstract}

%
%
\section{Introduction}
\label{sec:introduction}

Empirical risk minimization (ERM) is a central tool in supervised machine learning that enables the characterization, among others, of sample complexity and probably approximately correct (PAC) learning in a wide range of settings~\cite{vapnik1992principles}.
The application of ERM in the study of theoretical guarantees spans related disciplines such as machine learning \cite{vapnik1964perceptron}, information theory \cite{rodrigues2021information,mezard2009information} and statistics \cite{wainwright2019high, vershynin2018high}.
Classical problems such as classification\cite{blumer1989learnability,guyon1991structural}, pattern recognition \cite{lugosi1995nonparametric,bartlett1998sample}, regression \cite{vapnik1993local, cherkassky1999model}, and density estimation \cite{lugosi1995nonparametric,vapnik1999overview} can be posed as special cases of the ERM problem \cite{vapnik1999overview,bottou2018optimization}.
Unfortunately, ERM is prone to training data memorization, a phenomenon also known as overfitting \cite{krzyzak1996nonparametric,deng2009regularized,arpit2017Memorization}. For that reason, regularization is used to bound the sensitivity of the solution model to training data and provide generalization guarantees \cite{bousquet2002stability,vapnik2015uniform,aminian2021exact}.
Regularization establishes a preference over the models by encoding features of interest that conform to prior knowledge.

In different statistical learning frameworks, such as Bayesian learning \cite{robert2007bayesian,mcallester1998pacBayesian} and PAC learning \cite{valiant1984theory,shawe1997pac,cullina2018pac}, the prior knowledge over the set of models can be described by a reference probability measure.
Nonetheless, more general references can be adapted as proved in \cite{Perlaza-ISIT-2022} for the case of $\sigma$-finite measures.
In either case, the solution to the ERM problem can be cast as a probability distribution over all the candidate models.
A common regularizer is the relative entropy of the solution with respect to the reference over the set of models \cite{vapnik1999overview,raginsky2016information,russo2019much,zou2009generalization}.
The resulting problem formulation, termed ERM with relative entropy regularization (ERM-RER) has been extensively studied and its unique solution is the Gibbs probability measure, for which the most salient properties are well understood \cite{raginsky2016information,russo2019much,zou2009generalization,Perlaza-ISIT-2022,aminian2022information,Perlaza-ISIT2023b}.
Despite the many merits of the ERM-RER formulation, it has some significant limitations.
Firstly, the definition of the relative entropy in terms of the Radon-Nikodym derivative of the solution with respect to the reference probability measure, sets a hard barrier to the exploration of models outside the support of the reference. These models are not given any consideration by the resulting Gibbs probability measure regardless of the evidence provided by the training dataset.
Secondly, the choice of relative entropy over the alternatives often follows arguments based on upper bounds on the performance, which are hard to obtain and are not always informative when evaluated in practical settings \cite{wang2004enhancing,lin2014accelerated,yang2022estimation}.
For these reasons, exploring the asymmetry of the relative entropy is of particular interest to advancing the understanding of entropy regularization and its role in generalization.

Interestingly, there is no literature discussing the asymmetry of relative entropy in the context of ERM regularization. 
Hence, the issue of regularizing the ERM problem with the relative entropy of the reference with respect to the solution is an open problem. 
To differentiate between the two cases, we denote by Type-I the use of the relative entropy of the solution with respect to the reference;
and by Type-II the use of the relative entropy of the reference with respect to the solution.
This paper presents the solution to the Type-II ERM-RER problem and establishes a link to the Type-I ERM-RER problem via a transformation of the risk that can be cast as a tunable loss function  \cite{liao2018tunable, sypherd2019tunable, kurri2021realizing}.

The remainder of the paper is organized as follows. Section~\ref{sec:ERMproblem} presents the standard ERM problem.
Section~\ref{sec:Type1ERM_RER} describes the Type-I regularization.
The main contribution of the paper is the solution to the Type-II ERM-RER  presented in Section~\ref{sec:Type2ERM_RER}.
%
%
Section~\ref{sec:logERM_RER} studies the equivalence between Type-I  and Type-II regularization.
%
%
The conclusions are summarized in Section~\ref{sec:FinalRemarks}.

%
%
\section{Empirical Risk Minimization Problem}
\label{sec:ERMproblem}
The elements of the learning problem of interest are the sets \emph{models}, \emph{patterns}, and \emph{labels} denoted by $\set{M} \subseteq \reals^{d}$ with \mbox{$d \in \ints$}, $\set{X}$, and $\set{Y}$, respectively. 
A pair $(x,y) \in \mathcal{X} \times \mathcal{Y}$ is referred to as a \emph{labeled pattern} or \emph{data point}.
Several data points denoted by $(x_1, y_1)$, $( x_2, y_2 )$, $\ldots$, $( x_n, y_n )$ with $n \in \ints$, form a \emph{dataset}, which is represented by the tuple  $((x_1, y_1), (x_2, y_2), \ldots,(x_n, y_n))\in ( \set{X} \times \set{Y} )^n$.

Let the function~$f: \set{M} \times \mathcal{X} \rightarrow \mathcal{Y}$  be such that the label assigned to a pattern $x$ according to the model $\thetav \in \set{M}$ is $f(\thetav,x)$.
Then, given a dataset, the objective is to obtain a model $\thetav \in \set{M}$, such that, for all patterns $x \in \set{X}$, the  assigned label $f(\thetav,x)$ minimizes a notion of loss or risk. 
Let the function
\begin{equation}
\label{EqRiskFunDef}
    \ell: \set{Y} \times \set{Y} \rightarrow [0, +\infty),
\end{equation}
be such that given a data point $(x,y) \in \set{X} \times \set{Y}$, the loss or risk induced by choosing the model $\thetav \in \set{M}$ is $\ell(f(\thetav,x),y)$.
The risk function $\ell$ is assumed to be nonnegative and satisfy $\ell( y , y ) = 0$ for all $y\in\set{Y}$. 
Nonetheless, there might exist other models $\vect{\theta} \in \set{M}$ such that $\ell( f(\vect{\theta}, x'), y') = 0$ for the labelled data point $(x',y')$, revealing the need for a large number of labeled patterns for model selection.

The \emph{empirical risk} induced by a model $\vect{\theta}$ with respect to the dataset 
%
\begin{equation}
\label{EqTheDataSet}
\vect{z} = \big((x_1, y_1), (x_2, y_2 ), \ldots, (x_n, y_n )\big)  \in ( \set{X} \times \set{Y} )^n,
\end{equation}  
with $n \in \ints$, is determined by the function $\mathsf{L}_{\vect{z}}\!:\! \set{M} \rightarrow [0, +\infty)$, which satisfies
\begin{IEEEeqnarray}{rcl}
\label{EqLxy}
\mathsf{L}_{\vect{z}} (\vect{\theta} )  & \triangleq & 
\frac{1}{n}\sum_{i=1}^{n}  \ell ( f(\vect{\theta}, x_i), y_i ).
\end{IEEEeqnarray}
The ERM problem is given by the optimization problem
\begin{equation}
\label{EqOPfunction}
\min_{\vect{\theta} \in \set{M}} \mathsf{L}_{\vect{z}}  (\vect{\theta}  ),
\end{equation}
and the set of solutions to the problem is denoted by
\begin{equation}
\label{EqHatTheta}
\set{T} ( \vect{z}  ) \triangleq \arg\min_{\vect{\theta} \in \set{M}}    \mathsf{L}_{\vect{z}}  (\vect{\theta}  ).
\end{equation}
%
Note that if the set $\set{M}$ is finite, the ERM problem in~\eqref{EqOPfunction} has a solution, and therefore, it holds that $\abs{\set{T}(\dset{z})}>0$.
Nevertheless, in general, the ERM problem does not always have a solution;
that is, there exist choices of the loss function $\ell$ and the dataset $\dset{z}$ that yield $\abs{\set{T}(\dset{z})}=0$.
%
%

\subsection{Statistical Learning}
The Bayesian and PAC frameworks in \cite{shawe1997pac} and \cite{mcallester1998pacBayesian} solve the problem by constructing probability measures $P_{\Thetam|\vect{Z}=\dset{z}}$ conditioned on the dataset $\dset{z}$, from which models are randomly sampled.
In this context, finding probability measures that are minimizers of the ERM problem in~\eqref{EqOPfunction} over the set $\bigtriangleup\msblspc{\set{M}}$ of all probability measures that can be defined on the measurable space~$\msblspc{\set{M}}$, requires a metric that enables assessing the goodness of the probability measure.
A common metric is the notion of expected empirical risk.

\begin{definition}[Expected Empirical Risk]
\label{DefEmpiricalRisk}
Given a dataset \mbox{$\dset{z} \in  ( \set{X} \times \set{Y} )^n$}, 
let  the function $\mathsf{R}_{\dset{z}}: \triangle\msblspc{\set{M}} \rightarrow  [0, +\infty  )$~be such that for all probability measures $Q \in \triangle \msblspc{\set{M}}$,
\begin{equation}
\label{EqRxy}
\foo{R}_{\dset{z}}( Q ) \triangleq \int \foo{L}_{ \dset{z} } ( \thetav )  \diff Q(\thetav),
\end{equation}
where the dataset $\dset{z}$ is defined in~\eqref{EqTheDataSet};
and the function $\foo{L}_{\dset{z}}$ is defined in~\eqref{EqLxy}.
\end{definition}

The expected empirical risk is an important performance indicator of learning algorithms. However, it only gives an indication of the risk induced over the training dataset, while the performance of the ERM solutions is characterized by their generalization capability and sensitivity \cite{zou2009generalization,russo2019much,Perlaza-ISIT-2022,Perlaza-ISIT2023b}.
In the following, we review the Type-I relative entropy regularization that serves as the basis for the analysis of the regularization asymmetry.

%

%
%
\section{The Type-I ERM-RER Problem}
\label{sec:Type1ERM_RER}
The Type-I ERM-RER problem is parametrized by a probability measure $Q \in \triangle \msblspc{\set{M}}$ and a positive real $\lambda$, where the measure $Q$ is the \emph{reference measure} and $\lambda$ is the  \emph{regularization factor}. 
The {Type-I} ERM-RER problem, with parameters~$Q$ and~$\lambda$, consists of the following optimization problem:
%
\begin{IEEEeqnarray}{rCl}
\label{EqOpType1ERMRERNormal}
    \min_{P \in \bigtriangleup_{Q}\msblspc{\set{M}} } &\ & \foo{R}_{\dset{z}} ( P )  + \lambda \KL{P}{Q},
\end{IEEEeqnarray}
%
where the dataset~$\dset{z}$ is defined in~\eqref{EqLxy}, the function~$\foo{R}_{\dset{z}}$ is defined in~\eqref{EqRxy}, and the optimization domain~is 
\begin{IEEEeqnarray}{rCl}
\label{DefSetTriangUp}
		\bigtriangleup_{Q}\msblspc{\set{M}}   & \triangleq & \{P\in \bigtriangleup\msblspc{\set{M}}: P\ll Q \},
\end{IEEEeqnarray}
where the notation~$P\ll Q$ stands for~$P$ being absolutely continuous with respect to~$Q$.

%
The solution to the Type-I ERM-RER problem in~\eqref{EqOpType1ERMRERNormal} is the Gibbs probability measure \cite{russo2019much,raginsky2016information,Perlaza-ISIT-2022}, which is presented by the following lemma.

%
\begin{lemma}[\text{\cite[Lemma 1]{Perlaza-ISIT-2022}}]
\label{LemmOptimalModelType1}
Given a probability  measure $Q \in \bigtriangleup\msblspc{\set{M}}$ and a dataset $\dset{z} \in (\set{X} \times \set{Y})^n$, let the function $K_{Q,\dset{z}}: \reals \rightarrow \reals$ be such that for all $t \in \reals$,
\begin{equation}\label{EqDefKfunction}
		K_{Q,\dset{z}}( t ) = \log (\int \exp(t\foo{L}_{\dset{z}}(\thetav))\diff Q(\thetav)),
\end{equation}
where the dataset~$\dset{z}$ is defined in~\eqref{EqTheDataSet}.
Let also the set $\set{K}_{Q,\dset{z}} \subseteq \reals$ be
\begin{equation}
\label{EqDefSetK}
	\set{K}_{Q,\dset{z}} \triangleq \{s > 0: K_{Q,\dset{z}}(-\frac{1}{s}) < +\infty \}.
\end{equation}
Then, for all~$\thetav \in \supp Q$ and for all~$\lambda \in \set{K}_{Q,\dset{z}}$, the solution of the Type-I ERM-RER problem in~\eqref{EqOpType1ERMRERNormal}, is the unique probability measure~$\Pgibbs{P}{Q}\!\in\! \triangle_{Q}\msblspc{\set{M}}$, whose \RadonNikodym derivative with respect to~$Q$  satisfies~that
\begin{equation}
\label{EqGenpdfType1}
\frac{\diff \Pgibbs{P}{Q}}{\diff Q} ( \thetav ) = \exp(-K_{Q,\dset{z}}(-\frac{1}{\lambda})-\frac{1}{\lambda}\foo{L}_{\dset{z}}(\thetav)).
\end{equation}
\end{lemma} 

%
%
\section{The Type-II ERM-RER Problem}
\label{sec:Type2ERM_RER}

The Type-II ERM-RER problem is parametrized by a probability measure $Q \in \bigtriangleup\msblspc{\set{M}}$ and a positive real $\lambda$.
The measure $Q$ is referred to as the \emph{reference measure} and $\lambda$ as the \emph{regularization factor}.
Given the dataset~$\dset{z} \in (\set{X} \times \set{Y})^n$ in~\eqref{EqTheDataSet}, the Type-II ERM-RER problem, with parameters~$Q$ and~$\lambda$, consists of the following optimization problem:
\begin{IEEEeqnarray}{rcl}
    \min_{P \in \bigtriangledown_{Q}\msblspc{\set{M}} } &\ & \foo{R}_{\dset{z}} ( P )  + \lambda \KL{Q}{P},\label{EqOpType2ERMRERNormal}
\end{IEEEeqnarray}
where~$\dset{z}$ is defined in~\eqref{EqLxy}, the function~$\foo{R}_{\dset{z}}$ is defined in~\eqref{EqRxy}, and the optimization domain~is 
\begin{IEEEeqnarray}{rCl}
\label{DefSetTriangDown}
		\bigtriangledown_{Q}\msblspc{\set{M}} & \triangleq & \{P\in \bigtriangleup\msblspc{\set{M}}: Q\ll P \}.
\end{IEEEeqnarray}

%
%
\subsection{The Solution to the Type-II ERM-RER Problem}

The asymmetry of the relative entropy poses a distinct challenge when tackling the optimization problem given in~\eqref{EqOpType2ERMRERNormal}. The approach that leads to the solution of the Type-I in~\eqref{EqOpType1ERMRERNormal} needs to be adapted to accommodate the challenges posed by the absolute continuity requirement in \eqref{DefSetTriangDown}.
%
%
The solution of the Type-II ERM-RER problem in~\eqref{EqOpType2ERMRERNormal} is presented in the following theorem.
%

\begin{theorem}
\label{Theo_ERMType2RadNikMutualAbs}
Given a measure $Q \in \bigtriangleup \msblspc{\set{M}}$ and a dataset $\dset{z} \in (\set{X} \times \set{Y})^n$, let the function $\bar{K}_{Q,\dset{z}}: \reals \rightarrow \reals$ be such that for all $t \in (0, \infty)$ it holds that
\begin{equation}
\label{EqType2Krescaling}
	\bar{K}_{Q,\dset{z}}(t) = \beta,
\end{equation}
where
\begin{equation}
\label{EqType2KrescConstrain}
	\int \frac{t}{\beta + \foo{L}_{\dset{z}}(\thetav)} \diff Q(\thetav) = 1,
\end{equation}
with~$\foo{L}_{\dset{z}}$ being the function defined in~\eqref{EqLxy}.
The function~$\bar{K}_{Q,\dset{z}}$ in~\eqref{EqType2Krescaling} is well defined for a subset of ~$(0, \infty)$, which is denoted by~$\set{\bar{K}}_{Q,\dset{z}}$, and satisfies 
\begin{equation}
\label{EqDefSetKType2}
	\bar{\set{K}}_{Q,\dset{z}} \triangleq \{ t \in (0, \infty):\int \frac{t}{\bar{K}_{Q,\dset{z}}(t) + \foo{L}_{\dset{z}}(\thetav)} \diff Q(\thetav) = 1\}.
\end{equation}
%

Then, for all $\thetav \in \supp Q$ and for all $\lambda \in \bar{\set{K}}_{Q,\dset{z}}$, the solution to the optimization problem in \eqref{EqOpType2ERMRERNormal} is the unique probability measure $\Pgibbs{\bar{P}}{Q}$, whose \RadonNikodym derivative with respect to the probability measure $Q$ satisfies 
\begin{equation}
\label{EqGenpdfType2}
\frac{\diff \Pgibbs{\bar{P}}{Q}}{\diff Q} ( \thetav ) =  \frac{\lambda}{\bar{K}_{Q,\dset{z}}(\lambda) + \foo{L}_{\dset{z}}(\thetav)},
\end{equation}
where the functions $\mathsf{L}_{\vect{z}}$ and $\bar{K}_{Q,\dset{z}}$ are defined in~\eqref{EqLxy} and  \eqref{EqType2Krescaling}, respectively. 

\end{theorem}
\begin{IEEEproof}
	The proof is divided into two parts.
	In the first part, an ancillary optimization problem is solved in a subset of the optimization domain of the Type-II ERM-RER problem.
	%
	In the second part, it is shown that the solution obtained in this subset is, in fact, the solution of the Type-II ERM-RER problem.

	The first part is as follows.
	Given the dataset $\dset{z} \in (\set{X} \times \set{Y})^n$ in~\eqref{EqTheDataSet}, the ancillary optimization problem is given by:
	\begin{IEEEeqnarray}{rcl}
	\label{EqOpType2ERMRERancillary}
	\min_{P \in \bigcirc_{Q}\msblspc{\set{M}} } &\ & \foo{R}_{\dset{z}} ( P )  + \lambda \KL{Q}{P},
	\end{IEEEeqnarray}
	where the optimization domain is
	\begin{equation}
	\label{Eq_ProofThT2_setofmeasures}
		\bigcirc_Q \msblspc{\set{M}} \triangleq \bigtriangledown_{Q}\msblspc{\set{M}} \cap \bigtriangleup_{Q}\msblspc{\set{M}},
	\end{equation}
	and the sets~$\bigtriangleup_Q\msblspc{\set{M}}$ and~$\bigtriangledown_Q\msblspc{\set{M}}$ are respectively defined in~\eqref{DefSetTriangUp} and \eqref{DefSetTriangDown}.
	The solution to the ancillary optimization problem 	in~\eqref{EqOpType2ERMRERancillary} is presented by the following lemma.
	
	\begin{lemma}
	\label{lemm_Type2RNDbigcirc}
	For all~$\lambda \in \bar{\set{K}}_{Q,\dset{z}}$ with $\bar{\set{K}}_{Q,\dset{z}}$ in \eqref{EqType2Krescaling}, the solution to the optimization problem in~\eqref{EqOpType2ERMRERancillary} is the unique probability measure~$\Pgibbs{\bar{P}}{Q}$ in~\eqref{EqGenpdfType2}.
	\end{lemma}
	\begin{IEEEproof}
	From the fact that, for all $P\in\bigcirc_{Q}\msblspc{\set{M}}$, the measure $Q$ is mutually absolute continuous with respect to $P$, the ancillary optimization problem in~\eqref{EqOpType2ERMRERancillary} can be written as follows:
	\begin{IEEEeqnarray}{rCl}
	\label{EqType2ERMRERNormal}
	    \min_{P \in \bigcirc_{Q}\msblspc{\set{M}} } & & [\ \int \foo{L}_{\vect{z}}(\thetav)\frac{\diff P}{\diff Q}(\thetav)\diff Q(\thetav) \right.\nonumber \\
	    & & \quad -\> \left. \lambda \int \log(\frac{\diff P}{\diff Q}(\thetav))\diff Q(\thetav)],\qquad \\
	    \textrm{s.t.} \qquad \  &\ & \int \frac{\diff P}{\diff Q}(\thetav) \diff Q(\thetav) = 1.
	    \label{EqType2ConstTxt}
	\end{IEEEeqnarray}

	The Lagrangian of the optimization problem in~\eqref{EqType2ERMRERNormal} can be constructed in terms of a function in the set $\mathscr{M}$ of nonnegative measurable functions with respect to the measurable spaces $\bigcirc_Q \msblspc{\set{M}}$ and $\msblspc{\reals}$.
	%
	Let $L: \mathscr{M}\times \reals \rightarrow \reals$ be the Langragian
	\begin{IEEEeqnarray}{rCl}
	\label{EqFunctionalAllpf}
	   L (\frac{\diff P}{\diff Q},\beta) 
	   & = \int (\foo{L}_{\vect{z}}(\thetav)\frac{\diff P}{\diff Q}(\thetav)- \lambda \log(\frac{\diff P}{\diff Q}(\thetav))\right. \nonumber\\
	   & +\>\left. \beta( \frac{\diff P}{\diff Q}(\thetav) -1))\diff Q(\thetav),
	\end{IEEEeqnarray}
	where $\beta$ is a real value that acts as a Lagrange multiplier due to~\eqref{EqType2ConstTxt}.
	The Gateaux differential \cite{luenberger1997bookOptimization} of the functional $L$ in~\eqref{EqFunctionalAllpf} at $\left(\frac{\mathrm{d} P}{\mathrm{d} Q}, \beta\right) \in \mathscr{M}\times \reals$ in the direction of $h\in\mathscr{M}$ is
	\begin{IEEEeqnarray}{rCl}
	\label{EqGateauxDiffType2Pf}
    \partial L(\frac{\diff P}{\diff Q}, \beta; h )
    & = & \int h(\thetav)(\foo{L}_{\dset{z}}(\thetav)+\vphantom{\frac{\diff Q}{\diff P}}\beta \right.\nonumber\\
    &   & \> \left. -\lambda (\frac{\diff P}{\diff Q}(\thetav))^{-1} )\,\diff Q(\thetav).
	\end{IEEEeqnarray}
	The relevance of the Gateaux differential in~\eqref{EqGateauxDiffType2Pf} stems from \cite[Theorem $1$, page $178$]{luenberger1997bookOptimization}, which unveils the fact that a necessary condition for the functional $L$ in~\eqref{EqFunctionalAllpf} to have a stationary point at $( \frac{\diff  \Pgibbs{\bar{P}}{Q}}{\diff Q}, \beta ) \in \set{M} \times \reals$ is that for all functions $h \in \mathscr{M}$, the following holds:
	\begin{equation}
	\label{EqConditionhiType2}
	\partial L(\frac{\diff \Pgibbs{\bar{P}}{Q}}{\diff Q}, \beta; h)  = 0.    
	\end{equation}
	
	From the fact that $h$ is nonnegative, for all~$\thetav \in \set{M}$ it follows that
	\begin{equation}
	\label{Eq_ProofT2SemiLemma}
	\foo{L}_{\dset{z}}(\thetav) - \lambda(\frac{\diff \Pgibbs{\bar{P}}{Q}}{\diff Q}(\thetav))^{-1} + \beta  = 0,
	\end{equation}
	and thus,
	\begin{equation}
	\label{EqKlambInPsolverPf}
	  \frac{\diff \Pgibbs{\bar{P}}{Q}}{\diff Q}(\thetav) = \frac{\lambda}{\bar{K}_{Q,\dset{z}}(\lambda) + \foo{L}_{\dset{z}}(\thetav)},
	\end{equation}
	where the function $\bar{K}_{Q,\dset{z}}$ is defined in \eqref{EqType2Krescaling}.

	Finally, note that the objective function in~\eqref{EqType2ERMRERNormal} is the sum of two terms. 
	The first one, i.e., $ \int \foo{L}_{\dset{z}}( \thetav ) \frac{\diff P}{\diff Q}( \thetav )  \diff Q ( \thetav )$, is linear in  $\frac{\diff P}{\diff Q}$.
	The second, i.e., $-\int  \log (\frac{\diff P}{\diff Q}(\thetav)) \diff Q(\thetav)$, is strictly convex with  $\frac{\mathrm{d}P}{\mathrm{d}Q}$.
	Hence, given that $\lambda>0$, the sum of both terms is strictly convex with  $\frac{\diff P}{\diff Q}$.  This implies the uniqueness of $\Pgibbs{\bar{P}}{Q}$.
	\end{IEEEproof}

This completes the first part of the proof of Theorem~\ref{Theo_ERMType2RadNikMutualAbs}. The second part rests in the following lemma. 
%
	\begin{lemma}
	\label{lemm_RadNikDevMutualIneq}
		For all~$\lambda \in \bar{\set{K}}_{Q,\dset{z}}$, with $\bar{\set{K}}_{Q,\dset{z}}$ in \eqref{EqDefSetKType2}, it holds that
		\begin{equation}
		\label{EqNonTrievialAncillaryIneq}
		\min_{P \in \bigtriangledown_Q\setminus\bigcirc_Q } \foo{R}_{\dset{z}}(P) + \lambda \KL{Q}{P} > \min_{P \in \bigtriangledown_Q } \foo{R}_{\dset{z}}(P) + \lambda \KL{Q}{P}.
		\end{equation}
	\end{lemma}
	\begin{IEEEproof}
		The proof is presented in \cite{InriaRR9508}.
	\end{IEEEproof}

More specifically, Lemma~\ref{lemm_RadNikDevMutualIneq} conveys the fact that the relative entropy regularization penalty for considering models outside of the support is always greater than the reduction in the expected empirical risk induced by including these models. This includes the case in which the set $\set{T}(\dset{z})$ in~\eqref{EqHatTheta} lies outside of the support of $Q$. 


	From~\eqref{Eq_ProofThT2_setofmeasures}, it holds that
	\begin{equation}
	\label{Eq_ProofT2supset}
		\bigcirc_{Q}\msblspc{\set{M}} \subseteq \ \bigtriangledown_{Q}\msblspc{\set{M}}.
	\end{equation}
	Hence, from~\eqref{Eq_ProofT2supset}, it follows that
	\begin{equation}
	\min_{P \in \bigtriangledown_{Q}}
	\foo{R}_{\dset{z}}(P) + \lambda \KL{Q}{P} \leq \min_{P \in \bigcirc_{Q}} \foo{R}_{\dset{z}}(P) + \lambda \KL{Q}{P}.
	\label{Eq_ProofT2IneOp}
	\end{equation}
	From Lemma~\ref{lemm_RadNikDevMutualIneq}, it holds that
	\begin{equation}
	\label{Eq_ProofT2IneOp2}
		\min_{P \in \bigtriangledown_{Q}} \foo{R}_{\dset{z}}(P) + \lambda \KL{Q}{P}
		\geq \min_{P \in \bigcirc_{Q}} \foo{R}_{\dset{z}}(P) + \lambda \KL{Q}{P}.
	\end{equation}
	Thus, the measure $\Pgibbs{\bar{P}}{Q}$ in~\eqref{EqGenpdfType2} is the solution of the optimization problem in~\eqref{EqOpType2ERMRERNormal}, which completes the proof of Theorem~\ref{Theo_ERMType2RadNikMutualAbs}. 
\end{IEEEproof}

\subsection{Properties of the Solution}

The properties of the function $\bar{K}_{Q,\dset{z}}$ in~\eqref{EqType2Krescaling} and the set $\bar{\set{K}}_{Q,\dset{z}}$ in~\eqref{EqDefSetKType2} can be studied using the following mathematical objects. Given a positive real~$\delta$ and the dataset~$\dset{z}$ in~\eqref{EqTheDataSet}, consider the set
\begin{equation}
\label{EqType2LsetLamb2zero}
	\set{L}_{\dset{z}}(\delta) \triangleq \{\thetav \in \set{M}: \foo{L}_{\dset{z}}(\thetav) \leq \delta \},
\end{equation}
where the function~$\foo{L}_{\dset{z}}$ is defined in~\eqref{EqLxy} and $\delta \in [0,\infty)$.
Consider also the nonnegative real
\begin{equation}
\label{EqDefDeltaStar}
	\delta^\star_{Q,\dset{z}} \triangleq \inf \{\delta \in [0, \infty): Q(\set{L}_{\dset{z}}(\delta))>0\},
\end{equation}
with~$Q$ in~\eqref{EqOpType2ERMRERNormal}.
Let also $\set{L}^{\star}_{Q,\dset{z}}$ be the following level set of the empirical risk function $\foo{L}_{\dset{z}}$ in \eqref{EqLxy}:
\begin{equation}
\label{EqDefSetLStarQz}
	\set{L}^{\star}_{Q,\dset{z}} \triangleq \{\thetav \in \set{M}: \foo{L}_{\dset{z}}(\thetav) = \delta^\star_{Q,\dset{z}}\}.
\end{equation}

The following lemma introduces the properties of the function $\bar{K}_{Q,\dset{z}}$ in~\eqref{EqType2Krescaling}.

\begin{lemma}
\label{lemm_InfDevKtype2}
The function~$\bar{K}_{Q,\dset{z}}$ in~\eqref{EqType2Krescaling}, for fixed~$Q$ and~$\dset{z}$, is strictly increasing, continuous, and differentiable infinitely many times.
\end{lemma}
\begin{IEEEproof}
		The proof is presented in \cite{InriaRR9508}.
\end{IEEEproof}

Note that from Lemma~\ref{lemm_InfDevKtype2}, the value $\bar{K}_{Q,\dset{z}}(\lambda)$ in~\eqref{EqType2Krescaling} increases as the regularization factor $\lambda$ increases, which is consistent with the notion that it acts as a scaling factor in~\eqref{EqGenpdfType2}. This highlights its dependence with the dataset $\dset{z}$ in~\eqref{EqTheDataSet} and the reference measure $Q$ in~\eqref{EqOpType2ERMRERNormal}.

Similarly, the set $\bar{\set{K}}_{Q,\dset{z}}$ in~\eqref{EqDefSetKType2} also depends on the dataset~$\dset{z}$ in \eqref{EqTheDataSet} and the probability measure $Q$~in~\eqref{EqOpType2ERMRERNormal}.
The following lemma presents the properties of the set~$\bar{\set{K}}_{Q,\dset{z}}$ in~\eqref{EqDefSetKType2}.

\begin{lemma}
\label{lemm_Type2_kset}
	The set~$\bar{\set{K}}_{Q,\dset{z}}$ in~\eqref{EqDefSetKType2} is either the empty set or the set
	\begin{subequations}
	\label{Eq_Type2KsetsubFull}
	\begin{equation}
	\label{Eq_Type2Ksetsub}
		\bar{\set{K}}_{Q,\dset{z}} = (0,\infty).
	\end{equation}
	Moreover, for all $\lambda \in \bar{\set{K}}_{Q,\dset{z}}$ it holds that
	\begin{equation}
	\label{Eq_Type2KConstrain}
		\bar{K}_{Q,\dset{z}}(\lambda) \in (-\delta^\star_{Q,\dset{z}},\infty),
	\end{equation}
	\end{subequations}
	with $\bar{K}_{Q,\dset{z}}$ defined in \eqref{EqType2Krescaling} and $\delta^\star_{Q,\dset{z}}$ in \eqref{EqDefDeltaStar}. 
\end{lemma}
\begin{IEEEproof}
	The proof is presented in \cite{InriaRR9508}.
\end{IEEEproof}

Lemma~\ref{lemm_T2PropertiesEmpRiskSolution} below shows that the expected empirical risk induced by the Type-II ERM-RER solution can be computed in terms of the regularization factor $\lambda$ and the function $\bar{K}_{Q,\dset{z}}$ defined in~\eqref{EqDefSetKType2}. 
The relation of the expected empirical risk induced by $\Pgibbs{\bar{P}}{Q}$ in~\eqref{EqGenpdfType2} is presented by the following lemma.

\begin{lemma}
\label{lemm_T2PropertiesEmpRiskSolution}
	For all~$\lambda \in \bar{\set{K}}_{Q,\dset{z}}$, with $\bar{\set{K}}_{Q,\dset{z}}$ in~\eqref{EqDefSetKType2}, it holds that
	\begin{equation}
	\label{EqCharExEmpRiskT2}
		\foo{R}_{\dset{z}}(\Pgibbs{\bar{P}}{Q}) = \lambda - \bar{K}_{Q,\dset{z}}(\lambda),
	\end{equation}
	where the functions~$\foo{R}_{\dset{z}}$ and~$\bar{K}_{Q,\dset{z}}$ are respectively defined in~\eqref{EqRxy} and \eqref{EqType2Krescaling};
	and the measure~$\Pgibbs{\bar{P}}{Q}$ is defined in~\eqref{EqGenpdfType2}.
\end{lemma}
\begin{IEEEproof}
	The proof is presented in \cite{InriaRR9508}.
\end{IEEEproof}

%
The equality in~\eqref{EqCharExEmpRiskT2} provides an upper bound to the expected empirical risk $\foo{R}_{\dset{z}}(\Pgibbs{\bar{P}}{Q})$. 
The following corollary of Lemma~\ref{lemm_T2PropertiesEmpRiskSolution} formalizes this observation.

\begin{corollary}
\label{coro_boundtype2lambda}
	For all $\lambda\in \bar{\set{K}}_{Q,\dset{z}}$, with $\bar{\set{K}}_{Q,\dset{z}}$ in~\eqref{EqDefSetKType2}, it holds that
	\begin{IEEEeqnarray}{rCl}
		\foo{R}_{\dset{z}}(\Pgibbs{\bar{P}}{Q}) & < & \lambda + \delta^\star_{Q,\dset{z}}, 
	\end{IEEEeqnarray}
	where $\Pgibbs{\bar{P}}{Q}$ is the probability measure in \eqref{EqGenpdfType2} and  $\delta^\star_{Q,\dset{z}}$ is defined in \eqref{EqDefDeltaStar}.
\end{corollary}
The upper bound presented in Corollary~\ref{coro_boundtype2lambda} is useful as it gives operational meaning to the regularization factor.
Indeed, this bound shows that the regularization factor governs
the expected empirical risk increase with respect to the infimum of the empirical risk over the support.


\subsection{Discussion on Regularization Properties}

The Type-II relative entropy regularizer for the ERM problem in \eqref{EqOpType2ERMRERNormal} allows for an exploratory minimization, \ie models outside the support of the reference measure are given consideration.
However, Theorem~\ref{Theo_ERMType2RadNikMutualAbs} shows that the support of the probability measure $\Pgibbs{\bar{P}}{Q}$ in~\eqref{EqGenpdfType2} collapses into the support of the reference.
A parallel can be established  between Type-I and Type-II, as in both cases the support of the solution is the support of the reference measure.
In a nutshell, the use of relative entropy regularization inadvertently forces the solution to coincide with the support of the reference regardless of the training data.

%
%
\section{Interplay Between the Relative Entropy Asymmetry and the Risk}
\label{sec:logERM_RER}

This section presents a connection between the Type-I ERM-RER in~\eqref{EqOpType1ERMRERNormal} and Type-II ERM-RER problems in~\eqref{EqOpType2ERMRERNormal}. 
The log empirical risk is the function $\foo{V}_{\dset{z},\lambda}:\set{M} \rightarrow \reals$, which satisfies

%
\begin{equation}
\label{EqLogLxy}
	\foo{V}_{\dset{z},\lambda}(\thetav) \triangleq \log(\bar{K}_{Q,\dset{z}}(\lambda) + \foo{L}_{\dset{z}}(\thetav)),
\end{equation}
where the functions $\foo{L}_{\dset{z}}$ and $\bar{K}_{Q,\dset{z}}$ are defined in~\eqref{EqLxy} and~\eqref{EqType2Krescaling}, respectively. 
For the case in which $\bar{\set{K}}_{Q,\dset{z}}\neq\emptyset$, 
replacing the empirical risk in~\eqref{EqOPfunction} by the notion of log empirical risk in~\eqref{EqLogLxy} leads to the \emph{expected log empirical risk}, as shown hereunder.

\begin{definition}[Expected Log Empirical Risk]
\label{DefLogEmpiricalRisk}
Given a dataset $\dset{z} \in  ( \set{X} \times \set{Y} )^n$, 
let  the function $\mathsf{\bar{R}}_{\dset{z}}: \triangle\msblspc{\set{M}}  \rightarrow \reals$ be such that for all probability measures $P \in \triangle \msblspc{\set{M}}$ and for all $\lambda \in (0,+\infty)$ it holds that
\begin{equation}
\label{EqLogRxy}
\bar{\foo{R}}_{\dset{z},\lambda}( P ) \triangleq \int \foo{V}_{\dset{z},\lambda}(\thetav)  \diff P(\thetav),
\end{equation}
where the function $\foo{V}_{\dset{z},\lambda}$ is defined in~\eqref{EqLogLxy}.
\end{definition}
By considering the expected log empirical risk, an alternative formulation of the Type-I ERM-RER problem is presented.
This formulation, also parametrized by $Q$ and $\lambda$, consists in the following optimization problem: 
\begin{IEEEeqnarray}{rcl}
\label{EqOpERMLinkT1_T2}
    \min_{P \in \bigtriangleup_{Q}\msblspc{\set{M}}} & \quad & \bar{\foo{R}}_{\dset{z},\lambda}( P ) + \KL{P}{Q}.
\end{IEEEeqnarray}
Using the elements above, the main result of this section is presented in the following theorem.
%
\begin{theorem}
\label{Theo_ERMlogType1To2}
The solution to the optimization problem in~\eqref{EqOpERMLinkT1_T2} is the unique probability measure $\Pgibbs{\bar{P}}{Q}$ in~\eqref{EqGenpdfType2}.
\end{theorem}
%
\begin{IEEEproof}
Denote by $\Pgibbs{\hat{P}}{Q}$ the solution to the optimization problem in \eqref{EqOpERMLinkT1_T2}. Then, from Lemma~\ref{LemmOptimalModelType1}, for all $\thetav\in \supp{Q}$, it follows that
\begin{subequations}
\label{EqProoT1ToT2}
\begin{IEEEeqnarray}{rcl}
	\frac{\diff \Pgibbs{\hat{P}}{Q}}{\diff Q}(\thetav) 
	& = & \frac{\exp(-\foo{V}_{\dset{z},\lambda}(\thetav))}{\int \exp(-\foo{V}_{\dset{z},\lambda}(\nuv))\diff Q(\nuv)}  
	\label{EqProoT1ToT2_s1}\\
	& = & \frac{\exp(\log(\frac{1}{\foo{L}_{\dset{z}}(\thetav) + \bar{K}_{Q,\dset{z}}(\lambda)}))}{\int\! \exp(\log(\!\frac{1}{\foo{L}_{\dset{z}}(\nuv) + \bar{K}_{Q,\dset{z}}(\lambda)}\!)\!)\!\diff Q(\nuv)} 
	\label{EqProoT1ToT2_s2}\\
	& = & \frac{(\int \frac{1}{\foo{L}_{\dset{z}}(\nuv) + \bar{K}_{Q,\dset{z}}(\lambda)}\!\diff Q(\nuv))^{-1}}{\foo{L}_{\dset{z}}(\thetav)\! + \! \bar{K}_{Q,\dset{z}}(\lambda)}
	\label{EqProoT1ToT2_s3}\\
	& = & \frac{\lambda}{\foo{L}_{\dset{z}}(\thetav) + \bar{K}_{Q,\dset{z}}(\lambda)}
	\label{EqProoT1ToT2_s4}\\
	& = & \frac{\diff \Pgibbs{\bar{P}}{Q}}{\diff Q}(\thetav)
	 \label{EqProoT1ToT2_s5},
\end{IEEEeqnarray}
\end{subequations}
where equality~\eqref{EqProoT1ToT2_s2} follows from the definition of log empirical risk in~\eqref{EqLogLxy};
equality~\eqref{EqProoT1ToT2_s4} follows from~\eqref{EqType2Krescaling} and~\eqref{EqType2KrescConstrain};
and equality~\eqref{EqProoT1ToT2_s5} follows from Theorem~\ref{Theo_ERMType2RadNikMutualAbs}, which completes the proof.
\end{IEEEproof}
Theorem~\ref{Theo_ERMlogType1To2} establishes an equivalence between Type-I and Type-II regularization. It is shown therein that the direction of the relative entropy regularizer can be switched by appropriately transforming the risk function as shown in \eqref{EqLogLxy}.
Indeed, solving the Type-I ERM-RER problem with the expected log empirical risk defined in \eqref{EqLogRxy} yields the probability measure $\Pgibbs{\bar{P}}{Q}$ that is the solution to the Type-II ERM-RER problem.
In view of this, it is not surprising that the support for the probability measure that is the solution to the Type-II ERM-RER collapses into the support of the reference measure. In fact, the mutual absolute continuity between the solution and the reference probability measures is a consequence of the relative entropy regularization, regardless of its direction. 
Type-I regularization forces the support of the solution to include all the models in the support of the reference measure; on the other hand, Type-II regularization constrains the models in the support of the solution to the models in the support of the reference measure.


%
%
\section{Final Remarks}
\label{sec:FinalRemarks}

This work has introduced the Type-II ERM-RER problem and has presented its solution through Theorem~\ref{Theo_ERMType2RadNikMutualAbs}.
The solution highlights that regardless of the direction in which relative entropy is used as a regularizer, the models that are considered by the solution are necessarily in the support of the reference measure. In that sense, the restriction over the models introduced by the reference measure cannot be bypassed by the training data when relative entropy is used as the regularizer.
We have shown that this is a consequence of the equivalence that can be established between Type-I and Type-II regularization. 
Remarkably, the direction of the relative entropy regularizer can be switched by a logarithmic transformation of the risk. The mutual absolute continuity of both Type-I and Type-II ERM-RER solutions relative to the reference measure can be understood in the light of the equivalence between both types of regularization. 
The analytical results have also enabled us to provide an operationally meaningful characterization of the expected empirical risk induced by the Type-II solution in terms of the regularization parameters. This is turn reduces the computational burden of bounding the expected empirical risk. 
Moreover, the insight provided by the bounds on the expected empirical risk can be distilled into guidelines for the selection of the regularization parameter.


%
\IEEEtriggeratref{23}
\bibliographystyle{IEEEtran}
\bibliography{ISITBibStyle.bib}
%

\newpage

 \appendices



\end{document}